\documentclass[aps,pre,twocolumn,showpacs,floatfix]{revtex4}
\usepackage{amsmath,bm,epsfig}
\usepackage{latexsym}
\usepackage{amssymb}
\usepackage{color}
\usepackage{morefloats}
\usepackage{subfigure}
\usepackage{graphicx}
\usepackage{amsmath}
\usepackage{bm}
\usepackage{epsf,epsfig,graphics}
\usepackage{float}
\usepackage{makeidx}
\usepackage{multirow}

\begin{document}

\title{Effects of hidden nodes on the reconstruction of bidirectional networks}
\author{Emily S.C. Ching\footnote{ching@phy.cuhk.edu.hk}
and P.H. Tam} \affiliation{Department of Physics, The Chinese
University of Hong Kong, Shatin, Hong Kong}

\date{\today}
\begin{abstract}
Much research effort has been devoted to developing methods for
reconstructing the links of a network from dynamics of its nodes.
Many current methods require the measurements of the dynamics of
all the nodes be known. In real-world problems, it is common that
either some nodes of a network of interest are unknown or the
measurements of some nodes are unavailable. These nodes, either
unknown or whose measurements are unavailable, are called hidden
nodes. In this paper, we derive analytical results that explain
the effects of hidden nodes on the reconstruction of bidirectional
networks. These theoretical results and their implications are
verified by numerical studies.

\end{abstract}

 \maketitle

\maketitle

\section{Introduction}

Many systems of interest in physics and biology are represented by
complex networks of a large number of elementary units or nodes
that interact or link with each other~\cite{Strogatz}. A
substantial amount of data have been obtained for various networks
especially biological networks, and a grand challenge is to reveal
the structure of these networks, namely the links, their direction
and relative coupling strength, from the measured data. It is
expected that~\cite{Timme0} the structure of a network controls
its dynamics and thus one might be able to uncover information
about the structure of a network from its dynamics. Much research
effort has been devoted to developing methods for reconstructing a
network from the dynamics of the nodes~(see e.g.
\cite{TimmeReview,PhysReportYCLai2016} for review).
Counterintuitively, it has been demonstrated that the presence of
noise acting on the network can be beneficial for network
reconstruction as the noise induces a relation between measurable
quantities from dynamics and the network structure~\cite{noise}.
Making use of different relations of this kind, a number of
methods~\cite{PRE,PRErapid,SolvingPRE,PRErapid2,LaiPRE,ChenEPL,correlated}
have been proposed for reconstructing networks solely from the
dynamics of the nodes. In all these methods, in order to calculate
the quantities that are related to the network structure, the
measurements of the dynamics of all the nodes are required.

In real-world problems, it is common that either some nodes of a
network of interest are unknown or the measurements of some nodes
are unavailable. These nodes, either unknown or whose measurements
are unavailable, are called hidden nodes. It is thus important to
study and understand the effects of hidden nodes on the
reconstruction of
networks~\cite{DunnPRE2013,YCLai2014,ECC2014,WXWangPRL2015,HuangJPhyA2015,HuGang2017,Marc2017}.
This task is highly challenging and, as of today, there is not yet
a general and analytical understanding of the effects of hidden
nodes.

A usual practice infers links from the correlation of the
measurements, with a larger correlation coefficient interpreted as
a higher probability of
link~\cite{StuartScience2003,BrainPRL,frontiers}. However,
correlation between measurements of two nodes cannot be equated
with direct interactions between the two nodes. In fact, it has
been clearly shown that for networks of neurons, the spiking
dynamics of neurons can have weak pairwise correlations even
though they are strongly coupled~\cite{weak}. This study further
shows that the spiking dynamics of neurons are quantitatively
captured by the probability distribution $P_{\rm
Ising}(\hat\sigma_1, \ldots, \hat \sigma_N) \propto \exp [
\sum_{i<j}^N J_{ij} \hat \sigma_i \hat \sigma_j + \sum_i h_i \hat
\sigma_i ]$ of an Ising model, which is the maximum entropy
distribution of a system of $N$ nodes with binary state variables
$\hat \sigma_i= \pm 1$ that is consistent with the measured
averages and covariances. This leads to extensive studies of the
inverse Ising problem: the inference of the couplings $J_{ij}$,
which are interpreted as effective interactions between nodes $i$
and $j$, from measured averages and covariances of the data~(see
e.g.~\cite{Advance} for review).

In general, systems of interest have nodes with continuous state
variables. In this case, the maximum entropy distribution
consistent with the measured averages and covariances is the
multivariate Gaussian distribution~\cite{Conrad}
\begin{equation}
P_G(x_1, \ldots, x_N) = \frac{\exp\left[-({\bm x}- {\bm m})^T {\bm
\Sigma}^{-1} ({\bm x}- {\bm m})/2\right]}{(2\pi)^{N/2} \sqrt{{\rm
det}({\bm \Sigma})}}
 \label{multiGaussian}
\end{equation}
where $\bm{x} = (x_1, x_2, \ldots, x_N)^T$, $\bm m=(m_1, \ldots,
m_N)^T$ with $m_i$ being the measured average of $x_i$,
$\bm{\Sigma}$ is the measured covariance matrix, and the
superscript $T$ denotes the transpose.
Equation~(\ref{multiGaussian}) suggests that information of links
in bidirectional networks is contained in $\bm{\Sigma}^{-1}$, the
inverse of the covariance matrix. A mathematical relation between
the weighted adjacency matrix of the network and the inverse of
the covariance matrix has indeed been
derived~\cite{PRErapid,correlated} for a model class of
bidirectional networks with diffusive-like coupling and subjected
to Gaussian white noise. Based on this theoretical result, a
method that reconstructs bidirectional networks from
$\bm{\Sigma}^{-1}$ has been developed~\cite{correlated}.

In this paper, we address the question of how hidden nodes affect
the reconstruction of bidirectional networks. Using the same model
class of bidirectional networks studied
in~\cite{PRErapid,correlated}, we derive analytical results
relating quantities involving the hidden nodes to the inverse of
the covariance matrix of the measured data and use these results
to explain the effects of hidden nodes on the reconstruction
results. Then we carry out numerical studies to verify these
theoretical results and their implications.

\section{Formulation of the problem}

We consider weighted bidirectional networks of $N$ nodes with
nonlinear dynamics and diffusive-like coupling. The dynamics of
each node is described by a variable $x_i(t)$, $i=1,2,\ldots, N$,
and the time evolution of $x_i(t)$ is given by
\begin{equation}
\frac{d x_{i}} {dt} = f(x_i) + \sum_{j\ne i} g_{ij} A_{ij}
h(x_j-x_i) + \eta_i .\ \label{network}
\end{equation}
The adjacency matrix element $A_{ij}$ is
$1$ when node $j$ is linked to node $i$ by the diffusive-like coupling function $h$ with coupling strength
$g_{ij}$; otherwise $A_{ij} = g_{ij} = 0$. The coupling is
bidirectional so $A_{ij}=A_{ji}$ and $g_{ij}=g_{ji}$, and the
graph of the networks has no self-loops, that is,
 $A_{ii} \equiv 0$. As discussed~\cite{PRErapid,correlated}, $f$ describes the intrinsic dynamics that is
generally nonlinear and identical for all the nodes, and the
diffusive-like coupling function $h$
 satisfies $h(-z) = -h(z)$ and $h'(0)>0$.
 Thus excitatory or activating links have with $g_{ij} >0$ whereas inhibitory links have $g_{ij}<0$. Here we take $h'(0)=1$.
External influences are modelled by a Gaussian white noise $\eta$
with zero mean and variance $\sigma_n^2$:
\begin{equation} \overline{\eta_i(t)\eta_j(t')} = \sigma_n^2
\delta_{ij} \delta(t-t') \ , \label{white} \end{equation} where
the overbar is an ensemble average over different realizations of
the noise.

The weighted Laplacian matrix of the network, ${\bm L}$, is given by
\begin{equation}
    {L}_{ij} = s_{i} \delta_{ij} - g_{ij} A_{ij} \ , \quad s_{i} \equiv \sum_{k=1}^N g_{ik} A_{ik}
\end{equation}
and contains connectivity information of the network. Here $s_{i}$
is the weighted degree or the strength of node $i$. For these
networks, $x_i(t)$'s approach $X_0$ with $f'(X_0)<0$ in the
absence of noise.  Let $\delta x_i(t) = x_i(t) - X_0$, then the
linearized system around the noise-free steady state is given by
\begin{equation}
\frac{d}{dt} \delta x_i =  -\sum_j \left( L_{ij} + a \delta_{ij}
\right) \delta x_j + \eta_i  \label{deltax}
\end{equation}
where  $a \equiv - {f'(X_{0})}> 0$. We consider systems that have
stationary dynamics and this implies $\bm{L} + a{\bm I}$ is
positive
definite~\cite{Arnold}. Using Eq.~(\ref{deltax}),
 it has been derived~\cite{correlated} that the covariance
matrix $\bm{\Sigma}$, defined by
\begin{equation}
    \bm{\Sigma} \equiv \lim_{t \to \infty} \overline { [ \bm{x}(t) - \overline{ \bm{x} (t) } ] \, [ \bm{x}(t) - \overline{ \bm{x} (t) } ]^{T}}
\end{equation}
is related to $\bm{L}$  by
\begin{equation}
    \bm{\Sigma}^{-1} =  \frac{2}{\sigma_n^2} (\bm{L} + a  \bm{I}_N)  \label{oldresult}
\end{equation}
where ${\bm I}_N$ is the $N \times N$ identity matrix.
Equation~(\ref{oldresult}) is an exact result for the linearized
system~[Eq.~(\ref{deltax})] and a good approximation for the
original nonlinear network~[governed by Eq.~(\ref{network})] when
the noise is weak. All the theoretical results presented in this
paper should be understood in this manner.  An important
consequence of Eq.~(\ref{oldresult}) is
\begin{equation} \bm{\Sigma}^{-1}_{ij} = - \frac{2}{\sigma_n^2} g_{ij}
A_{ij}\ ,  \qquad i \ne j \label{conseq} \end{equation} which
indicates that the off-diagonal elements of $\Sigma^{-1}_{ij}$
would separate into two groups according to $A_{ij} = 0$ or 1.
Making use of this result, a reconstruction method of the
adjacency matrix and thus the links of the network  by performing
clustering analysis of the off-diagonal elements of
$\bm{\Sigma}^{-1}$ has been developed~\cite{correlated}. In this
method, ${\bm \Sigma}$ is evaluated by approximating the ensemble
average by time average:
\begin{equation}
    {\Sigma}_{ij} \approx \langle [x_i(t) - \langle x_i (t) \rangle][x_j(t) - \langle x_j(t) \rangle] \rangle
    \label{Sigmaapp}
\end{equation}
where $\langle \cdots \rangle$ denotes a time average. To evaluate
$\bm{\Sigma}^{-1}$, the measurements of $\bm{x}(t)$ from all the
$N$ nodes are required.

We study the problem of network reconstruction when there are
$n_h$ hidden nodes and only measurements from $n = N-n_h < N$
nodes are available. We call these $n$ nodes the measured nodes
and, for clarity, denote their measured dynamics by $y_i(t)$,
$i=1, 2, \ldots, n$ and the corresponding covariance matrix of the
measured data by ${\bm \Sigma}_m$:
\begin{equation}
    \bm{\Sigma}_m \equiv \lim_{t \to \infty} \overline { [ \bm{y}(t) - \overline{ \bm{y} (t) } ] \, [ \bm{y}(t) - \overline{ \bm{y} (t) }
    ]^{T}}
\end{equation}
where $\bm{y}(t) \equiv (y_{1},  y_{2}, \cdots , y_{n})^{T}$.
Similarly,
\begin{equation} (\Sigma_m)_{ij}
    \approx \langle [y_i(t) - \langle y_i (t) \rangle][y_j(t) - \langle y_j(t) \rangle] \rangle
    \end{equation}
We would like to answer the following questions: How would the
hidden nodes affect the reconstruction results based on ${\bm
\Sigma}_m^{-1}$? Whether and when the links among the $n$ measured
nodes can be reconstructed from the measured $y_i(t)$'s?

\section{Theoretical relation for ${\bm \Sigma}_m^{-1}$}

Without loss of generality, we let $y_i(t)=x_i(t)$, $i=1, 2,
\ldots, n$. Then we partition ${\bm \Sigma}$ into four block
matrices
\begin{align}
    \bm{\Sigma} =
    \begin{pmatrix}
    \bm{\Sigma}_{m} & \bm{U} \\
    \bm{U^{T}} & \bm{\Sigma}_h
    \end{pmatrix}
\end{align}
where the $n \times n$ block matrix is ${\bm \Sigma}_m$, the $n_h
\times n_h$ block matrix ${\bm \Sigma}_h$ is the covariance matrix
of the hidden nodes given by \begin{equation} (\Sigma_h)_{\mu \nu}
= \lim_{t \to \infty} \overline { [ x_{\mu+n}(t) - \overline{
x_{\mu+n} (t) } ] \, [ x_{\nu+n}(t) - \overline{ x_{\nu+n} (t) }
]} \label{Sigmah}    \end{equation} and the $n \times n_h$ block
matrix ${\bm U}$ measures the covariance between the measured and
hidden nodes with
\begin{equation}
U_{i\mu} =  \lim_{t \to \infty} \overline { [ y_i(t) - \overline{
y_{i} (t) } ] \, [ x_{\mu+n}(t) - \overline{ x_{\mu+n} (t) }  ]}
 \label{U}   \end{equation}
 For clarity, we use Roman subscripts $i,j, \ldots$ for the measured
 nodes and Greek subscripts $\mu, \nu, \ldots$ for the hidden
 nodes. We partition the weighted Laplacian matrix in a similar fashion:
\begin{align}
    \bm{{L}} =
    \begin{pmatrix}
    \bm{L}_{m} & \bm{E} \\
    \bm{E}^{T} & \bm{L}_h
    \end{pmatrix} \label{block}
\end{align}
The $n \times n$ block matrix ${\bm L}_m$ and the $n_h \times n_h$
block matrix ${\bm L}_h$, with elements $(L_m)_{ij} = L_{ij}$ and
$(L_h)_{\mu \nu} = L_{\mu+n,\nu+n}$, contain information of the
connectivity among the measured nodes and among the hidden nodes
respectively while the $n \times n_h$ block matrix ${\bm E}$
contains information of the connectivity between the measured and
hidden nodes with elements
\begin{equation}
E_{i\mu} = -g_{i,\mu+n} A_{i,\mu+n} \label{E} \end{equation}

Using Eq.~(\ref{oldresult}), we obtain
\begin{eqnarray}
     \bm{I}_n &=& \frac{2}{\sigma_n^2} \left[ \bm{\Sigma_m} \left({\bm L_m}+ a {\bm I}_n \right) + \bm{UE^{T}}  \right]\label{r1c1} \\
     \bm{0} &=&  \bm{\Sigma}_{m} {\bm E} + \bm{U}\left({\bm L}_h  + a \bm{I}_{n_h} \right) \label{r1c2}
\end{eqnarray}
which imply
\begin{equation}
    \bm{\Sigma}_{m}^{-1} = \frac{2}{\sigma_n^2} \left[\bm{L}_m + a \bm{I}_n  - \bm{E} \left(\bm{L}_h + a \bm{I}_{n_h} \right)^{-1} \bm {E}^T  \right]
    \label{newresult}
\end{equation}
As ${\bm L}+a{\bm I}_N$ is positive definite, ${\bm L}_h + a{\bm
I}_{n_h}$ is also positive definite and is thus invertible.
 We define
\begin{equation}
\bm{C} \equiv \bm{E} \left(\bm{L}_h + a \bm{I}_{n_h} \right)^{-1}
\bm {E}^T \label{corr}
\end{equation} then the off-diagonal elements of Eq.~(\ref{newresult})
can be written as
\begin{equation}
\left( \Sigma_m^{-1} \right)_{ij} = - \frac{2}{\sigma_n^2} \left(
g_{ij}  A_{ij} + C_{ij} \right) \ , \qquad i \ne j \label{newoff}
\end{equation} which shows that the hidden nodes affect the
reconstruction results based on ${\bm \Sigma}_m^{-1}$ by
introducing corrections given by $\bm C$. This can be seen
directly by using Eq.~(\ref{oldresult}) to rewrite
Eq.~(\ref{newresult}) as
\begin{equation}
    \left( \Sigma_m^{-1} \right)_{ij} = {\Sigma}^{-1}_{ij} - \frac{2}{\sigma_n^2} C_{ij}  \qquad i, j = 1, \ldots, n \label{newresult2}
\end{equation}
 Equations~(\ref{newresult}) and (\ref{newresult2}) are our major theoretical
 results and we shall use them to answer the questions of interest.

\section{Corrections due to hidden nodes}
\begin{figure}[htbp]
\centering \epsfig{file=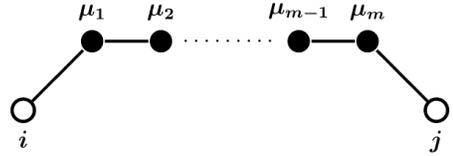,width=6cm} \caption{A path connecting the measured
nodes $i$ and $j$ (open circles) via the hidden nodes $\mu_1, \mu_2,
\ldots, \mu_{m-1}, \mu_m$~(closed circles).} \label{fig1}
\end{figure}

From Eq.~(\ref{corr}) and using Eq.~(\ref{E}), we immediately see
that $C_{ij}=0$ when $g_{i, \mu+n}=0$ or $g_{j,\mu+n}=0$ for all
$\mu=1, 2, \ldots, n_h$, that is when at least one of the measured nodes
$i$ and $j$ is not connected to any hidden node. We let $\bm{M}
\equiv \bm{L}_h + a \bm{I}_{n_h} \equiv {\bm S}-{\bm W}$ where
${\bm S}$ and ${\bm W}$ are defined by
\begin{eqnarray}
S_{\mu \nu} &=& (s_{\mu+n} +a )\delta_{\mu \nu} \label{S} \\
W_{\mu \nu} &=& g_{\mu+n, \nu+n} A_{\mu+n, \nu+n} \label{W}
\end{eqnarray}
for $\mu, \nu = 1, 2, \ldots, n_h$. $\bm S$ is a diagonal matrix
with the diagonal elements related to the strength of the hidden nodes and
${\bm W}$ is the weighted adjacency matrix of the hidden nodes.
Then we obtain~(see Appendix)
\begin{eqnarray}
\nonumber && C_{ij} =
\sum_{\mu_1=1}^{n_{h}} F_0 W_{i \mu_1}
W_{\mu_1 j} + \sum_{\mu_1,\mu_2=1}^{n_{h}} F_1 W_{i\mu_1} W_{\mu_1 \mu_2}W_{\mu_2j} \nonumber \\
    && + \sum_{k=3}^{n_h-1} \sum_{\mu_1, \cdots, \mu_{k+1}=1}^{n_{h}}
F_{k} W_{i\mu_1}  \left(\prod_{j=1}^{k} W_{\mu_j \mu_{j+1}}
\right) W_{\mu_{k+1} j}  \label{Cexp} \qquad
\end{eqnarray}
where $F_k$ generally depends on $s_{\mu_l+n} + a$, $l=2, \ldots,
k$ and the eigenvalues of $\bm{M}$. We
have also similarly defined $W_{i \mu} \equiv g_{i,\mu+n}
A_{i,\mu+n}$ for $i=1, 2, \ldots, n$, $\mu =1, 2, \ldots, n_h$ and
$W_{\mu i}=W_{i \mu}$. The product $W_{i\mu_1} W_{\mu_1
\mu_2} \cdots W_{\mu_{m-1} \mu_m} W_{\mu_m j}$ is nonzero only if
there is a path connecting the measured node $i$ and the measured
node $j$ via hidden nodes $\mu_1, \mu_2, \ldots, \mu_{m-1},
\mu_m$~(see Fig.~\ref{fig1}). Thus $C_{ij}$ would be zero if there does
not exist any path connecting the measured nodes $i$ and $j$ via
hidden nodes only. In general, we expect $C_{ij}$ to be nonzero
for some pairs of measured nodes $i$ and $j$.

As indicated by Eq.~(\ref{conseq}), the distribution of the
off-diagonal elements of $\Sigma^{-1}_{ij}$ for $i,j =1, 2, \ldots, n$ can be written as
\begin{equation}
P(\Sigma^{-1}_{ij} = x) = (1-r) P_0(x)+ r P_1(x)
\label{PSigma}
\end{equation}
where $r$ is the fraction of connected pairs among the measured
nodes and is equal to the number of connected pairs of measured
nodes divided by $n(n-1)/2$, $P_0$ and $P_1$ are the distributions
of $\Sigma^{-1}_{ij}$ with $A_{ij}=0$ and $A_{ij}=1$ respectively. If the
positive and negative coupling strength of the links are described
by two different distributions, $P_1$ can further be a mixture of
two distributions $P_{1+}$ and $P_{1-}$, which correspond to
$g_{ij}>0$ and $g_{ij}<0$ respectively~[c.f.
Eq.~(\ref{PSigma+-})]. In the limit of infinite number of data,
$P_0(x)$ approaches $\delta(x)$. For a finite number of data
points, numerical studies~\cite{correlated} show that $P_0(x)$ is
well-approximated by a Gaussian distribution of mean $m_0=0$ and
standard deviation $\sigma_0$, and $\sigma_0$ decreases when the
number of data points increases. Equation~(\ref{conseq}) implies
that $P_1(x)$ would have a mean $m_1$ and standard deviation
$\sigma_1$ given by
\begin{equation}
m_1 = - \frac{2}{\sigma_n^2} \langle g \rangle  \ ; \qquad
\sigma_1 = \frac{2}{\sigma_n^2} \sigma_g \label{P1stat}
\end{equation} where $\langle g \rangle$ and $\sigma_g$ are the average
and standard deviation of  the coupling strength $g_{ij}$ of the
links. In the presence of hidden nodes, the corrections $C_{ij}$
will modify the distributions $P_0$ and $P_1$ to ${\tilde P}_0$
and ${\tilde P}_1$:
\begin{equation}
P((\Sigma_m^{-1})_{ij}=x)= (1-r) {\tilde P}_0(x)+ r {\tilde
P}_1(x)\ , \ \ i \ne j \label{PSigmam}
\end{equation}
The mean and standard deviation of ${\tilde P}_i$, denoted by
$\tilde{m}_i$, $\tilde{\sigma}_i$, $i=0$ and  $1$, would be
modified. Using Eq.~(\ref{newoff}) and the results for $P_0(x)$,
we obtain
\begin{eqnarray} {\tilde m}_0
&=& -\frac{2}{\sigma_n^2} \mu_{C,0}
\label{est1} \\
{\tilde \sigma}_0^2 &\approx& \sigma_0^2 + \frac{4}{\sigma_n^4}
\sigma_{C,0}^2
\label{est2} \\
{\tilde m}_1 &=& m_1 -\frac{2}{\sigma_n^2} \mu_{C,1} \label{est3}
\\ {\tilde \sigma}_1^2 &=&  \sigma_1^2 + \frac{4}{\sigma_n^4}
\left[\sigma_{C,1}^2 + K(g_{ij},C_{ij}) \right] \label{est4} \ \
\end{eqnarray}
where
\begin{eqnarray} \mu_{C,0} & \equiv& \langle C_{ij} | A_{ij} =
0 \rangle \label{def1} \\
\mu_{C,1} &\equiv& \langle C_{ij} | A_{ij} =1 \rangle \label{def2} \\
\sigma_{C,0}^2 &\equiv& \langle C_{ij}^2 | A_{ij}=0 \rangle -
\langle
C_{ij} | A_{ij} = 0 \rangle^2 \label{def3} \\
\sigma_{C,1}^2 &\equiv& \langle C_{ij}^2 | A_{ij}=1 \rangle -
\langle
 C_{ij} | A_{ij} = 1 \rangle^2 \label{def4} \\
 K(g_{ij},C_{ij}) &\equiv& \langle g_{ij} C_{ij} | A_{ij} =1 \rangle
- \langle g \rangle \langle C_{ij} | A_{ij}=1 \rangle \qquad \label{def5}
\end{eqnarray}
Here $\mu_{C,i}$ and $\sigma_{C,i}$ for $i=1,2$ are the
conditional average and conditional standard deviation of $C_{ij}$
for $A_{ij}=0$ and $A_{ij}=1$ respectively, and $K(g_{ij},
C_{ij})$ measures the correlation of $C_{ij}$ with the coupling
strength $g_{ij}$ for measured nodes $i$ and $j$ that are
connected. Hence, the corrections $C_{ij}$
would shift the means, broaden  and distort the distributions of
$P_0$ and $P_1$ to $\tilde P_0$ and $\tilde P_1$.

Suppose $P_0$ and $P_1$ are distinguishable.  If ${\tilde P}_0$
and ${\tilde P}_1$ remain distinguishable even though with a
larger extent of overlap, then the links among the measured nodes
can still be reconstructed amid with a larger error rate. For a
mixture of two general distributions, there is no simple criterion
on when the component distributions are distinguishable.
Nonetheless, the component distributions are likely to be
distinguishable when the absolute value of the difference between
their means are larger than a certain multiple of the sum of their
standard deviations. Let $m_1-m_0=\gamma(\sigma_0+\sigma_1)$ with
$|\gamma|>1$. Using Eqs.~(\ref{est1})-(\ref{est4}), we obtain
\begin{eqnarray}
\nonumber && {\tilde m}_1-{\tilde m}_0 \\
\nonumber &=& \gamma \left \{({\tilde \sigma}_0^2 -\frac{4}{\sigma_n^2}{\sigma_{C,0}^2)^{1/2} +
[{\tilde \sigma}_1^2 -\frac{4}{\sigma_n^2}(\sigma_{C,1}^2+K)]^{1/2}}\right \}\\
&& \ \  - \frac{2}{\sigma_n^2}(\mu_{C,1}-\mu_{C,0})
\label{mod}
\end{eqnarray}
Thus ${\tilde P}_0$ and
${\tilde P}_1$ are likely to remain
 distinguishable (with $|{\tilde m}_1-{\tilde m}_0|/({\tilde \sigma}_0+{\tilde \sigma}_1)>1$) if $|\mu_{C,1}-\mu_{C,0}|$, $\sigma_{C,0}$,
 $\sigma_{C,1}$ and $|K(g_{ij},C_{ij})|$ are sufficiently small.

 To shed further light on this, we use Eq.~(\ref{corr}) to obtain a crude estimate of
$C_{ij}$.  Substitute ${\bm L}_h+a {\bm I}_{n_h} = {\bm S} -{\bm
W}= {\bm S}({\bm I}_{n_h}-{\bm S}^{-1} {\bm W})$ into
Eq.~(\ref{corr}), we have
\begin{equation}
 {\bm C} =
{\bm E}({\bm I}_{n_h} - {\bm S}^{-1} {\bm W})^{-1} {\bm S}^{-1}
{\bm E}^T \label{Nseries}
\end{equation}
If the Neumann series $\sum_{k=0}^\infty ({\bm S}^{-1}{\bm W})^k$
converges, it converges to $({\bm I}_{n_h} - {\bm S}^{-1} {\bm
W})^{-1}$. The necessary and sufficient condition for the Neumann
series to converge is the spectral radius of ${\bm S}^{-1}{\bm
W}$, denoted by $\rho({\bm S}^{-1}{\bm W})$, is less than
1~\cite{note}. For networks with $g_{ij} \ge 0$, one can easily
show that the infinity norm $||{\bm S}^{-1} {\bm W}||_\infty
\equiv \max_\mu \{\sum_\nu |({\bm S}^{-1} {\bm W})_{\mu \nu}| \}<
1$, and thus $\rho({\bm S}^{-1}{\bm W}) \le ||{\bm S}^{-1} {\bm
W}||_\infty <1$. For networks with both positive and negative
$g_{ij}$, we have checked numerically that $\rho({\bm S}^{-1}{\bm
W})<1$ for all the cases studied. When the Neumann series
converges, we keep two terms in the series, namely $({\bm I}_{n_h}
- {\bm S}^{-1} {\bm
 W})^{-1} \sim {\bm I}_{n_h} + {\bm S}^{-1}{\bm W}$, to
obtain a crude estimate of $C_{ij}$:
\begin{eqnarray}
\nonumber C_{ij} &\sim&
 \sum_{\mu=n+1}^{N} \frac{g_{i\mu} g_{j\mu} A_{i\mu} A_{j\mu}}{s_{\mu} +
 a}\\
 && + \sum_{\mu,\nu=n+1}^{N} \frac{g_{i\mu} g_{\mu\nu}g_{j\nu}A_{i\mu} A_{\mu\nu}A_{j\nu}}{(s_{\mu} + a)(s_{\nu}+a) } \label{Capprox}
    \end{eqnarray}
Using Eq.~(\ref{Capprox}), one sees that the magnitude of $C_{ij}$
depends on three factors: (1) the number of paths connecting the
measured nodes $i$ and $j$ via the hidden nodes which determines
the number of nonzero terms in the sums, (2) the strength of the
hidden nodes and (3) the coupling strength of the links in these
paths. Regarding to the second factor, we note that hidden nodes
with larger strength actually give rise to smaller corrections in
contrary to what one might have guessed. If these factors do not
differ much between the two groups of unconnected or connected
measured nodes, then $\mu_{C,1} \sim \mu_{C,0}$; if these factors
do not vary much among the measured nodes in each group, then
$\sigma_{C,0}$ and $\sigma_{C,1}$ would be small and if
 these factors do not correlate with the magnitude of $g_{ij}$
 for connected measured nodes, then $|K(g_{ij},C_{ij})|$ would be small
even when the magnitudes of the corrections $C_{ij}$'s themselves
might be large. Such situations are expected when the the hidden
nodes are not preferentially linked to the measured nodes in any
manner. In this case, ${\tilde P}_0$ and ${\tilde P}_1$ remain
distinguishable and it is possible to reconstruct the links among
the measured nodes from $(\Sigma_{m}^{-1})_{ij}$, $i \ne j$.

\section{Numerical results and discussions}

We check our theoretical results using data from numerical
simulations. We study five different networks, four of $N=100$
each and one of $N=1000$.
\begin{itemize}
\item[(1)] Network A: it consists of two random networks, each of
50 nodes and a connection probability of 0.2, connected to each
other by one link and $g_{ij}$'s, taken from a Gaussian
distribution $\mathcal N(10,2^2)$ of mean 10 and standard
deviation 2, are all positive. We take all the 50 nodes of one of
the random network as hidden nodes. \item[(2)] Network B: it is a
random network of connection probability 0.2 and $g_{ij}$'s also
taken from $\mathcal N(10,2^2)$ and are all positive. We choose
the hidden nodes randomly from the network with $n_h \le 70$ such
that the number of links among the measured nodes is at least of
the order of 100. \item[(3)] Network C: it is similar to network B
except that $g_{ij}$ of 80\% of the links taken from $\mathcal
N(10,2^2)$ and the remaining 20\% taken from $\mathcal
N(-10,2^2)$. As a result, about 80\% of the $g_{ij}$'s are
positive and about 20\% are negative. The hidden nodes are chosen
randomly from the network. \item[(4)] Network D: it is a
scale-free network of $N=1000$~\cite{BA} with
degree distribution obeying a power law and $g_{ij}$'s taken from
$\mathcal N(10,2^2)$ are all positive. The hidden nodes are chosen
randomly from the network. \item[(5)] Network E: it is constructed
by linking 30 additional nodes to a random network of $70$ nodes
and connection probability 0.2 with the restriction that every one
of the additional nodes is only commonly connected to randomly
selected pairs of unconnected nodes in the random network; and the
additional nodes are randomly connected among themselves with the
same connection probability 0.2. $g_{ij}$'s are taken from
$\mathcal N(10,2^2)$ and are all positive. We take the 30
additional nodes as hidden nodes.
\end{itemize}

For the dynamics, we mainly study nonlinear logistic function
\begin{equation}
f(x)= 10x(1-x) \label{logistic} \end{equation} and diffusive
coupling function
 \begin{equation}
h(y-x)=y-x \label{diffusive}
\end{equation} and take $\sigma_n=1$ for the noise.
To explore how general our theoretical results are, we go beyond the description by Eq.~(\ref{network}) and study two
additional cases. In the first
additional case, the nodes of network B have two-dimensional state
variables $(x_i(t),y_i(t))$ with
 nonlinear FitzHugh-Nagumo (FHN)
dynamics~\cite{FHN}
\begin{eqnarray} \label{FHN1}
\dot{x_i}&=&
(x_i-{x_i^3}/{3}-y_i)/\epsilon +\displaystyle\sum\limits_{j\ne i}
g_{ij} A_{ij}(x_j-x_i)+\eta_i \qquad
\\
\dot{y_i}&=&x_i+\alpha \,  \label{FHN2}
\end{eqnarray}
where $\epsilon=0.01$ and $\alpha=0.95$. In the second additional
case, the nodes of network B have three-dimensional state
variables $(x_i(t),y_i(t),z_i(t))$ with nonlinear R\"ossler
dynamics~\cite{Rossler} and nonlinear coupling~\cite{PRErapid}:
\begin{eqnarray}
\dot{x}_i
&=& -y_i -z_i  +\sum_{j\ne i} g_{ij}A_{ij} \tanh(x_j - x_i) +\eta_i \\
\dot{y}_i &=& x_i + c_1 y_i +\sum_{j\ne i} g_{ij}A_{ij}\tanh(y_j -
y_i)
\\
\dot{z}_i &=& c_2 + z_i(x_i-c_3) +\sum_{j\ne i}
g_{ij}A_{ij}\tanh(z_j - z_i) \  \qquad
\end{eqnarray}
where $c_1=c_2=0.2$ and $c_3=9$. In these two additional cases,
the system does not approach a steady state in the absence of
noise, and has chaotic dynamics when the nodes are decoupled in
the second case with R\"ossler dynamics. We integrate the
equations of motion using the Euler-Maruyama method and record the
time series $x_i(t)$ with a sampling interval $\delta t=5\times
10^{-4}$. For all the cases studied, including the cases with FHN
and R\"ossler dynamics, we calculate ${\bm \Sigma}$ using
$x_i(t)$'s with a time average over $N_{\rm data}=2 \times 10^6$
data points.

\begin{figure}[htbp]
\centering \epsfig{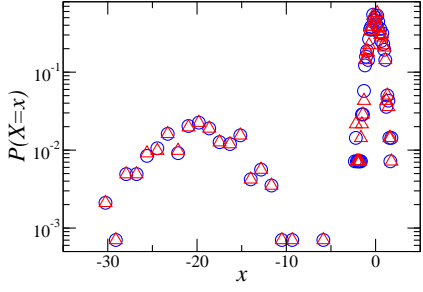} \caption{Comparison
of the distributions $P(X=x)$ of $X=\Sigma^{-1}_{ij}$~(circles)
and $X=(\Sigma_m^{-1})_{ij}$~(triangles) for $i \ne j = 1, 2,
\ldots, n=50$ for network A.} \label{fig2}
\end{figure}

For network A, since there is only one link connecting the hidden
nodes and the measured nodes, there is no path connecting any pair
of measured nodes via the hidden nodes thus $C_{ij}=0$ for all $i
\ne j$. As a result, Eq.~(\ref{newresult2})  implies that
$(\Sigma_m^{-1})_{ij} = \Sigma_{ij}^{-1}$ for $i \ne j$. We show
the distributions of $P(\Sigma_{ij}^{-1})$ and
$P((\Sigma_m^{-1})_{ij})$ for $i \ne j = 1, 2, \ldots, n=50$ in
Fig.~\ref{fig2}. As expected, the two distributions coincide with
each other. Moreover, $P(\Sigma_{ij}^{-1})$ is bimodal with the
peak around zero corresponding to $P_0$ for unconnected nodes and
the peak around $x_m \approx -20$ corresponding to $P_1$ for
connected nodes in accord with Eq.~(\ref{PSigma}). Furthermore,
the value of $x_m$ is in excellent agreement with the theoretical
value of $\mu_1=-2\langle g \rangle/\sigma_n^2$~[see
Eq.~(\ref{P1stat})]. Hence in this case, the links among the
measured nodes can be reconstructed from $(\Sigma_m^{-1})_{ij}$
with $i \ne j$, which can be calculated using the dynamics
$y_i(t)$ of the measured nodes only.

\begin{figure}[htbp]
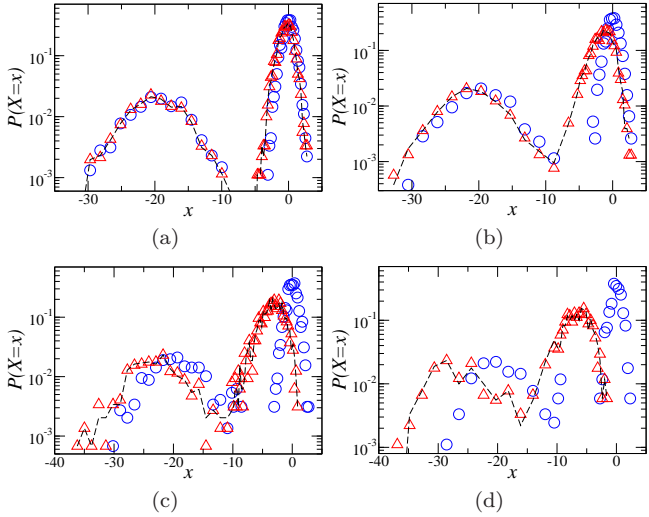

\centering \subfigure[]{\includegraphics*[width=4.2cm]{Fig3a.eps}}
\subfigure[]{\includegraphics*[width=4.2cm]{Fig3b.eps}}
\subfigure[]{\includegraphics*[width=4.2cm]{Fig3c.eps}}
\subfigure[]{\includegraphics*[width=4.2cm]{Fig3d.eps}}
\caption{Comparison of distributions $P(X=x)$ of
$X=\Sigma^{-1}_{ij}$~(circles) and
$X=(\Sigma_m^{-1})_{ij}$~(triangles) for network B with logistic
dynamics and different number of hidden nodes: (a) $n_h=10$, (b)
$n_h=30$, (c) $n_h=50$, and (d) $n_h=70$. The dashed lines are the
theoretical results of $X=\Sigma^{-1}_{ij} -2 C_{ij}/\sigma_n^2$
with $C_{ij}$ calculated using Eq.~(\ref{corr}).} \label{fig3}
\end{figure}

\begin{figure}[htbp]
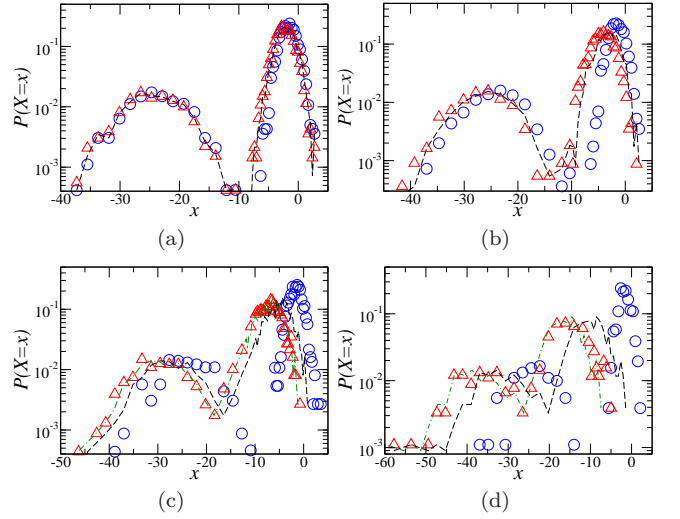

\centering \subfigure[]{\includegraphics*[width=4.2cm]{Fig4a.eps}}
\subfigure[]{\includegraphics*[width=4.2cm]{Fig4b.eps}}
\subfigure[]{\includegraphics*[width=4.2cm]{Fig4c.eps}}
\subfigure[]{\includegraphics*[width=4.2cm]{Fig4d.eps}}
\caption{Same as Fig.~\ref{fig3} for network B with FHN dynamics
but the dashed lines are now the theoretical estimates of
$X=\Sigma^{-1}_{ij} - (2/\sigma_n^2)(E L_h^{-1} E^T)_{ij}$. The
dot-dashed lines are the improved theoretical estimates of
$X=\Sigma^{-1}_{ij} - (2/\sigma_n^2)[(E L_h^{-1} E^T)_{ij} +b]$
with $b=1$ for (c) and $b=3$ for (d).} \label{fig4}
\end{figure}

\begin{figure}[htbp]
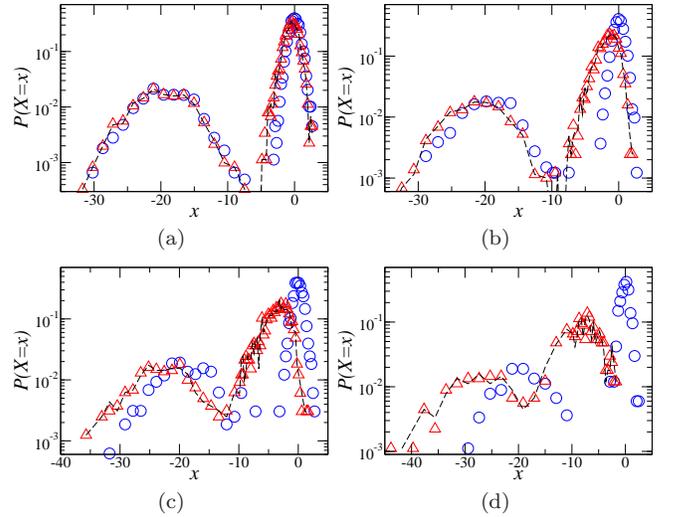

\centering \subfigure[]{\includegraphics*[width=4.2cm]{Fig5a.eps}}
\subfigure[]{\includegraphics*[width=4.2cm]{Fig5b.eps}}
\subfigure[]{\includegraphics*[width=4.2cm]{Fig5c.eps}}
\subfigure[]{\includegraphics*[width=4.2cm]{Fig5d.eps}}
\caption{Same as Fig.~\ref{fig4} for network B with R\"ossler
dynamics.} \label{fig5}
\end{figure}

For network B with the hidden nodes randomly chosen, there are
nonzero $C_{ij}$'s for some pairs of measured nodes $i$ and $j$.
We first consider the case with logistic dynamics and calculate
$C_{ij}$ using Eq.~(\ref{corr}) and together with
${\Sigma}^{-1}_{ij}$, we obtain the theoretical results for
$(\Sigma_m^{-1})_{ij}$ using Eq.~(\ref{newresult2}). We compare
these theoretical results with $(\Sigma_m^{-1})_{ij}$  directly
calculated from the measured dynamics $y_i(t)$'s in
Fig.~\ref{fig3} and perfect agreement is found for all the values
of $n_h$ studied. For FHN and R\"ossler dynamics, the system is
not described by Eq.~(\ref{network}) thus $a=-f'(X_0)$ is not
defined. We put $a=0$ in Eq.~(\ref{corr}) and obtain the
theoretical estimate for the off-diagonal $(\Sigma_m^{-1})_{ij}$
as $\Sigma^{-1}_{ij} - (2/\sigma_n^2)(E L_h^{-1} E^T)_{ij}$.
Interestingly, these theoretical estimates
 are in good agreement with the directly calculated
$(\Sigma_m^{-1})_{ij}$'s in most cases, as shown in
Figs.~\ref{fig4} and \ref{fig5}. For FHN dynamics with larger
$n_h$, an improved theoretical estimate is obtained by
$\Sigma^{-1}_{ij} - (2/\sigma_n^2)[(E L_h^{-1} E^T)_{ij} + b]$,
where $b$ is a constant. This indicates the general applicability
of our theoretical results beyond the model class studied.

As shown in Figs.~\ref{fig3}-\ref{fig5}, the distribution
$P((\Sigma_m^{-1})_{ij})$ is a mixture of the modified
distributions $\tilde P_0$ and $\tilde P_1$, in accord with
Eq.~(\ref{PSigmam}), which remain distinguishable as expected
since the hidden nodes are chosen randomly. Thus it is possible to
reconstruct the links among the measured nodes from
$(\Sigma_m^{-1})_{ij}$ with $i \ne j$. We note that this is true
for all the three kinds of dynamics studied and even when the
hidden nodes outnumber the measured nodes. In
Table~\ref{tab1}, we compare the error rates of the reconstruction
results obtained using k-means clustering of
$(\Sigma_m^{-1})_{ij}$ from the measured dynamics only and of
${\Sigma}^{-1}_{ij}$ from the dynamics of the whole network. We
measure the error rates by the ratios of false negatives (FN) and
false positives (FP) over the number of actual links $N_L$ among
the measured nodes. These error rates are related to the
sensitivity and specificity usually used for a predictive test:
sensitivity is given by $1 - {\rm FN}/N_L$ and specificity is
given by $1 - ({\rm FP}/N_L)\rho/(1 - \rho)$, where $\rho =
N_L/[n(n-1)/2]$ is the link density of the measured nodes.
For networks with low link density $\rho$, the error rates can be rather high even when specificity is close to 1 so
the error rates are better measures of the accuracy of the reconstruction results~\cite{PRErapid2}.  As can be seen, the accuracy of the
reconstruction results using $(\Sigma_m^{-1})_{ij}$ from the
measured nodes only is comparable to that obtained using
$(\Sigma^{-1})_{ij}$ from all the nodes.

\begin{table}[htp]
    \centering
    \begin{tabular}{|c|c|c|c|c|c|}
    \hline
        network & dynamics & \; $n_{h}$ \; & $\rho$ & FN/$N_{L}$ (\%) & FP/$N_{L}$ (\%) \\
    \hline
    A  & logistic & $50$ & 0.202 &  0.81 (0.81) & 0.00 (0.00)\\
    \hline
    \multirow{12}{*}{B} &  \multirow{4}{*}{logistic} &
     $10$ & \; 0.198 \; & $0.76 \; \; (0.88)$ & $0.00 \; \; (0.00)$\\
     & & $30$ &0.197& $1.05 \; \; (1.26)$ & $0.00 \; \; (0.00)$ \\
     & & $50$ & 0.202 &$1.62 \; \; (1.62)$ & $0.00 \; \; (0.00)$\\
     & & $70$ &0.218 & $1.05 \; \; (3.16)$ & $0.00 \; \; (0.00)$\\
    \cline{2-6}
    & \multirow{4}{*}{FHN} & $10$ & 0.198 & $0.63 \; \; (0.51)$ & $0.00 \; \; (0.00)$\\
     & & $30$ & 0.197 &$1.05 \; \; (0.84)$ & $0.00 \; \; (0.00)$ \\
     & & $50$ & 0.202 & $1.62 \; \; (1.21)$ & $0.00 \; \; (0.00)$\\
 & & $70$ & 0.218 & $3.16 \; \; (1.05)$ & $1.05 \; \; (0.00)$\\
    \cline{2-6}
     &\multirow{4}{*}{R\"{o}ssler} & $10$ & 0.198 & $1.01 \; \; (0.88)$ & $0.00 \; \; (0.00)$\\
     & & $30$ &0.197 & $1.10 \; \; (0.44)$ & $0.00 \; \; (0.00)$ \\
     & & $50$ & 0.202 & $0.87 \; \; (1.31)$ & $0.00 \; \; (0.00)$\\
     & & $70$& 0.218 & $3.57 \; \; (0.00)$ & $5.95 \; \; (0.00)$\\
    \hline
    \multirow{4}{*}{C} & \multirow{4}{*}{logistic} &
       $20$ & 0.198 & $0.96 \; \; (0.80)$ & $0.00 \; \; (0.00)$ \\
     & & $30$ & 0.197 & $0.85 \; \; (0.21)$ & $0.00 \; \; (0.00)$ \\
     & & $40$ & 0.199 &$0.85 \; \; (0.00)$ & $1.14 \; \; (0.00)$\\
     & & 60 & 0.206&  2.48 (0.00) & 1.86 (0.00) \\
    \hline
     \multirow{4}{*}{D} & \multirow{4}{*}{logistic} &
     100 &0.0039 &0.44 (0.50) &0.00 (0.00)\\
& &300 &0.0037 &0.45 (0.45) &0.11 (0.00)\\
& &500 &0.0040 &0.60 (0.60)& 1.20 (0.00)\\
& & 700& 0.0017 &1.05 (1.05)& 3.16 (0.00)\\
    \hline
     E &  logistic & $30$ &0.199 & $0.42 \; \; (0.42)$ & $29.88 \; (0.00)$\\
    \hline
    \end{tabular}
\caption{Error rates of the reconstruction results using k-means
clustering of $(\Sigma_{m}^{-1})_{ij}$ for the various networks
studied. $N=100$ for networks A, B, C and E and $N=1000$ for
network D. $\rho = N_L/[n(n-1)/2]$, where $N_L$ is the number of
links among the measured nodes and $n=N-n_h$ is the number of measured nodes. Two clusters
are used for all networks except network C where three clusters
are used. We show also the error rates using $\Sigma^{-1}_{ij}$ in
parentheses.} \label{tab1}
\end{table}

\begin{figure}[htbp]
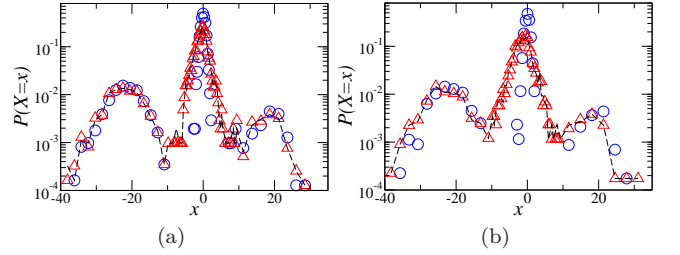

\centering \subfigure[]{\includegraphics*[width=4.2cm]{Fig6a.eps}}
\subfigure[]{\includegraphics*[width=4.2cm]{Fig6b.eps}}
\caption{Similar to Fig.~\ref{fig3} for network C with both
positive and negative $g_{ij}$'s for two different numbers of
hidden nodes: (a) $n_h=20$ and (b) $n_h=40$.} \label{fig6}
\end{figure}

For network C, the positive and negative $g_{ij}$'s follow two
different distributions so $P_1$ can be further decomposed into a
weighted sum of $P_{1+}$ and $P_{1-}$, which correspond to $g_{ij}
> 0$ and $g_{ij} <0$ respectively. Therefore, Eq.~(\ref{PSigma})
can be rewritten as
\begin{eqnarray}
&& P(\Sigma^{-1}_{ij} = x) \nonumber \\
 &=& (1-r) P_0(x) + r [\beta P_{1+}(x)+ (1-\beta)
P_{1-}(x)] \qquad \label{PSigma+-}
\end{eqnarray}
where $\beta$ is the fraction of the links among the measured
nodes with positive $g_{ij}$'s. Similarly, Eq.~(\ref{PSigmam}) is
also rewritten as
\begin{eqnarray}
&& P((\Sigma_m^{-1})_{ij} = x) \nonumber \\
&=& (1-r) \tilde P_0(x)+ r [\beta \tilde P_{1+}(x)+ (1-\beta)
\tilde P_{1-}(x)] \qquad \label{PSigmam+-}
\end{eqnarray}
As $g_{ij}$ can now be either positive or negative, the terms in
the summations contributing to $C_{ij}$~[see Eq.~(\ref{Capprox})]
could cancel one another. Thus we expect the magnitudes of
$\mu_{C,0}$,  $\mu_{C,1+} \equiv \langle C_{ij} | A_{ij}=1,
g_{ij}>0 \rangle$ and $\mu_{C,1-} \equiv \langle C_{ij} |
A_{ij}=1, g_{ij}<0 \rangle$ to be smaller than the magnitudes of
$\mu_{C,0}$ and $\mu_{C,1}$ for network B with logistic dynamics.
It can indeed be clearly seen that the shifts of $\tilde P_0$,
$\tilde P_{1+}$ and $\tilde P_{1-}$ from $P_0$, $P_{1+}$ and
$P_{1-}$ in Fig.~\ref{fig6} are smaller than the shifts of $\tilde
P_0$ and $\tilde P_{1}$ from $P_0$ and $P_{1}$ in Fig.~\ref{fig3}.
Moreover, $\tilde P_0$, $\tilde P_{1+}$ and $\tilde P_{1-}$ are
again only slightly broadened as compared with $P_0$, $P_{1+}$ and
$P_{1-}$ because the hidden nodes are randomly chosen. Thus for
network C, the links among the measured nodes can also be
accurately reconstructed from clustering of $(\Sigma_m^{-1})_{ij}$
with $i \ne j$~(see Table~\ref{tab1}).

\begin{figure}[htbp]
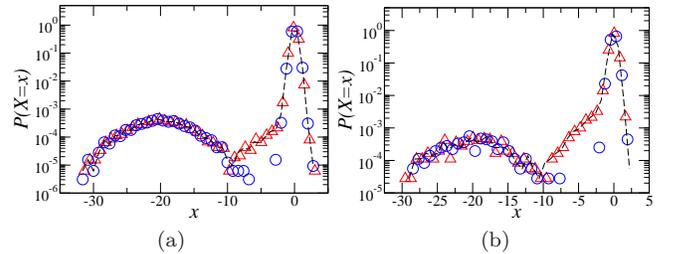

\centering
\subfigure[]{\includegraphics*[width=4.2cm]{newFig7a.eps}}
\subfigure[]{\includegraphics*[width=4.2cm]{newFig7b.eps}}
\caption{Similar to Fig.~\ref{fig3} for scale-free network D with
(a) $n_h=100$ and (b) $n_h=700$.} \label{fig7}
\end{figure}

For the scale-free network D, as the link density $\rho$ is very
small, most of the measured nodes are not linked via a path of
hidden nodes and thus most $C_{ij}$'s vanish. But as the nodes
have a power-law degree distribution, both the strength of the
hidden nodes and the number of paths connecting measured nodes via
hidden nodes could have a large variation leading to a large
variation in the magnitude of $C_{ij}$'s. This implies a large
$\sigma_{C,0}/\mu_{C,0}$ and $\sigma_{C,1}/\mu_{C,1}$ as compared
to the case of random network B. In particular, this results
in a larger distortion from $P_0$ to ${\tilde P}_0$ as seen in
Fig.~\ref{fig7}. The effect is more evident for $P_0$ because
there are far more unconnected measured nodes than connected
measured nodes due to the small $\rho$. Nonetheless, ${\tilde
P}_0$ and ${\tilde P}_1$ remain distinguishable and the error
rates of the reconstruction of the links among the measured nodes
remain low~(see Table~\ref{tab1}).

\begin{figure}[htbp]
\centering \epsfig{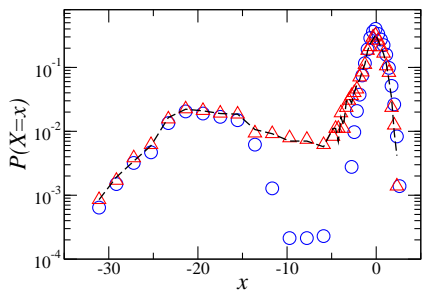}
\caption{Comparison of distributions $P(X=x)$ of
$X=\Sigma^{-1}_{ij}$~(circles) and
$X=(\Sigma_m^{-1})_{ij}$~(triangles) for network E with $n_h=30$
hidden nodes that are preferentially linked to measured nodes that
are unconnected. The dashed line is the theoretical result of
$X=\Sigma^{-1}_{ij} -2 C_{ij}/\sigma_n^2$ with $C_{ij}$ calculated
using Eq.~(\ref{corr}).} \label{fig8}
\end{figure}

In network E, every one of the $n_h=30$ hidden nodes is only
commonly linked to randomly selected pairs of unconnected measured
nodes.  We first randomly choose a pair of unconnected measured
nodes and link all the $n_h$ hidden nodes to both of them. Then we
link the hidden nodes to a second pair of unconnected measured
nodes with the restriction that no hidden nodes are commonly
linked to connected measured nodes that might exist among the
first and second pairs of unconnected nodes. Therefore, the number
of hidden nodes commonly linked to the second pair can be less
than $n_h$. We repeat the process for all the remaining pairs of
unconnected measured nodes. In this way, the number of hidden
nodes commonly linked to a given pair of measured nodes $i$ and
$j$ or the number of nonzero terms in the first sum in
Eq.~(\ref{Capprox}) is identically zero for $i$ and $j$ that are
connected, and varies
 among $i$ and $j$ that are unconnected. This preferential connection of the hidden nodes to unconnected measured nodes
gives rise to  $\mu_{C,0}
> \mu_{C,1}$ and $\sigma_{C,0} >\sigma_{C,1}$. Thus the distortion of $P_0$ is large leading to a larger overlap of $\tilde
P_0$ and $\tilde P_1$ as shown in Fig.~\ref{fig8}. As expected,
the error rates of the reconstruction results are larger with
FP/$N_L \approx 30$\%~(see Table~\ref{tab1}). However, we note
that even with this error rate, the specificity is above 90\%.
Moreover, the error rate FN/$N_L$ remains less than 1\% and thus
more than 99\% of the actual links among the measured nodes are
correctly reconstructed.

\vspace{0.5cm}

\section{Conclusions}

We have addressed the interesting question of how hidden nodes
affect reconstruction of bidirectional networks. By using a model
class of bidirectional networks with nonlinear dynamics and
diffusive-like coupling and subjected to a Gaussian white noise,
as described by Eq.~(\ref{network}), we have derived analytical
results, Eqs.~(\ref{newresult}) and Eq.~(\ref{newresult2}), that
allow us to answer this question precisely. Hidden nodes affect
the reconstruction results by introducing corrections $C_{ij}$.
 These corrections $C_{ij}$ are nonzero when
the measured nodes $i$ and $j$ are connected via a path of hidden
nodes as depicted in Fig.~\ref{fig1}. Our estimate of $C_{ij}$, as
shown in Eq.~(\ref{Capprox}), shows that three factors determine
$C_{ij}$'s, namely the number of paths of hidden nodes connecting
the two measured nodes $i$ and $j$, the coupling strength of the
links and the strength of the hidden nodes in these paths.
Interestingly, hidden nodes with larger strength give rise to
smaller corrections when the other two factors remain the same.
When the hidden nodes are not preferentially linked to the
measured nodes in any manner, these three factors would not differ
much between or among the two groups of connected and unconnected
measured nodes and, as a result, the hidden nodes would have
little effects on the reconstruction of the links among the
measured nodes. This is true even when the hidden nodes outnumber
the measured nodes. In the event that the hidden nodes are
preferentially linked to the measured nodes such that one or more
of the above three factors vary significantly either between or
among the two groups, the accuracy of the reconstruction results
would deteriorate. Yet useful information can still be uncovered.
We have verified our theoretical results and their implications
using numerical simulations and our numerical results indicate the
applicability of our results and analytical understanding beyond
the model class of bidirectional networks studied.

Hence our work shows that the method based on the inverse of
covariance is useful for reconstructing bidirectional networks
even when there are hidden nodes. Most networks of interest in
real-world problems are directed. It is highly challenging to
derive analytical results for the effects of hidden nodes on the
reconstruction of general directed networks. It would thus be
interesting to investigate whether and how the present results and
understanding for bidirectional networks can be extended to
general directed networks.

\begin{acknowledgments} The work of ESCC and PHT has been
supported by the Hong Kong Research Grants Council under grant no.
CUHK 14304017.
\end{acknowledgments}

\appendix
\section{Derivation of Eq.~(\ref{Cexp})}

Denote the eigenvalues of $\bm{M} \equiv \bm{L}_h + a {\bm
I}_{n_h}$ by $\lambda_k$, $k=1, \ldots, n_h$. By Cayley-Hamilton
theorem, $\bm{M}$ satisfies its own characteristic equation.
Therefore \begin{eqnarray} \nonumber \bm{0}
&=& \prod_{k=1}^{N-n} (\bm{M}-\lambda_k \bm{I}_{n_h})  \nonumber \\
&=& \bm{M}^{n_h} + \sum_{m=1}^{n_h} (-1)^m e_m \bm{M}^{n_h-m} \qquad \label{charEq}
\end{eqnarray} where ${\bm M}^0=\bm{I}_{n_h}$ and
$e_m(\lambda_1, \ldots, \lambda_k)$, $1\le m\le n_h$, are the
elementary symmetric polynomials of $\lambda_k$'s. For example,
$e_1= \sum_{k=1}^{n_h} \lambda_k$ and $e_{n_h}= \prod_{k=1}^{n_h}
\lambda_k$. Multiplying ${\bm M}^{-1}$ to Eq.~(\ref{charEq}) and
rearranging terms, we express ${\bm M}^{-1}$ as a finite
polynomial of $\bm M$:
\begin{eqnarray}
{\bm M}^{-1} &=&
\sum_{m=0}^{n_h-1} (-1)^{m} \frac{e_{n_h-1-m}}{e_{n_h}} \bm{M}^{m}
\qquad
\label{Minverse}
\end{eqnarray}

Using $\bm{M} \equiv \bm{S} - \bm{W}$ with $\bm{S}$ and $\bm{W}$
defined in Eqs.~(\ref{S}) and (\ref{W}), we obtain the elements of
${\bm M}^m$ in terms of the elements of $\bm S$ and $\bm W$. For
$m=1$ and $2$:
\begin{eqnarray}
  M_{\mu \nu} &=& f_0^{(1)} \delta_{\mu \nu} + f^{(1)}_1 W_{\mu \nu} \label{M1} \\
 (M^{2})_{\mu \nu}
 &=& f_0^{(2)} \delta_{\mu \nu} + f^{(2)}_1 W_{\mu \nu} + \sum_{\alpha = 1}^{n_{h}} f^{(2)}_2 W_{\mu \alpha} W_{\alpha \nu} \qquad
\label{M2}
\end{eqnarray}
and for $3 \le m \le n_h-1$:
\begin{eqnarray} && (M^{m})_{\mu \nu}
 =f^{(m)}_0 \delta_{\mu \nu} + f^{(m)}_{1} W_{\mu\nu} + \sum_{\alpha=1}^{n_h} f^{(m)}_2 W_{\mu \alpha} W_{\alpha \nu} \nonumber \\
 &&   + \sum_{k=3}^m \sum_{\alpha_1, \cdots, \alpha_{k-1}=1}^{n_{h}}  f^{(m)}_{k} W_{\mu \alpha_{1}} \left(\prod_{j=1}^{k-2} W_{\alpha_j\alpha_{j+1}} \right)W_{\alpha_{k-1} \nu}  \label{Mm} \qquad \ \
\end{eqnarray}
Here $f^{(m)}_0 = (s_{\mu+n}+a)^m$ and $f^{(m)}_m = (-1)^m$ for
$m=1, \ldots, n_h-1$. For $m \ge 2$, $f^{(m)}_k$, $1\le k \le
m-1$, generally depends on $\hat{s}_{\mu+n}  \equiv
{s}_{\mu+n}+a$, $\hat{s}_{\nu+n}$ and $\hat{s}_{\alpha_i+n}$,
$i=1, \ldots, k-1$. Explicit results for $f^{(m)}_{k}$, $1\le k
\le m-1$, for $m=2$ and 3 are
\begin{eqnarray}
  f^{(2)}_1 &=&  - (\hat{s}_{\mu+n} + \hat{s}_{\nu+n})
    \label{f21} \\
    f^{(3)}_1 &=&  - [ (\hat{s}_{\mu+n})^{2} + (\hat{s}_{\nu+n})^{2} + \hat{s}_{\mu+n} \hat{s}_{\nu+n}] \label{f31}\\
   f^{(3)}_2 &=&  \hat{s}_{\mu+n} + \hat{s}_{\nu+n} + \hat{s}_{\alpha+n}  \label{f32}
\end{eqnarray}

Substituting Eqs.~(\ref{M1})-(\ref{Mm}) into Eq.~(\ref{Minverse}),
we obtain
\begin{eqnarray}
&& (M^{-1})_{\mu \nu} = F_0 \delta_{\mu \nu} + F_1 W_{\mu \nu} +
\sum_{\alpha=1}^{n_h} F_2 W_{\mu \alpha} W_{\alpha \nu}\nonumber
\\
&& + \sum_{k=3}^{n_h-1} \sum_{\alpha_1, \cdots,
\alpha_{k-1}=1}^{n_h} F_k W_{\mu \alpha_1} \left(\prod_{j=1}^{k-2}
W_{\alpha_j\alpha_{j+1}} \right)W_{\alpha_{k-1} \nu} \qquad
\label{M-1ij}
\end{eqnarray}
where \begin{equation} F_k = \sum_{m=k}^{n_h-1} (-1)^m
\frac{e_{n_h-1-m}}{e_{n_h}} f_k^{(m)} , \ \ \ 0 \le k \le n_h-1 ,
\end{equation}
depends generally on $\hat{s}_{\mu+n}$, $\hat{s}_{\nu+n}$,
$\hat{s}_{\alpha_i+n}$, $i=1, \ldots, k-1$ and $\lambda_{k}$.
Putting Eq.~(\ref{M-1ij}) into Eq.~(\ref{corr}), we thus obtain
\begin{eqnarray}
  &&  C_{ij} = \sum_{\mu, \nu = 1}^{n_{h}} E_{i \mu} (M^{-1})_{\mu \nu} E_{j \nu} \nonumber \\
    &&= \sum_{\mu_1=1}^{n_{h}} F_0 W_{i \mu_1}
W_{\mu_1 j} + \sum_{\mu_1,\mu_2=1}^{n_{h}} F_1 W_{i\mu_1} W_{\mu_1 \mu_2}W_{\mu_2j} \nonumber \\
    && + \sum_{k=3}^{n_h-1} \sum_{\mu_1, \cdots, \mu_{k+1}=1}^{n_{h}}
F_{k} W_{i\mu_1}  \left(\prod_{j=1}^{k} W_{\mu_j\mu_{j+1}}
\right) W_{\mu_{k+1} j}  \label{dresult} \qquad \qquad
\end{eqnarray}
which is just Eq.~(\ref{Cexp}).

\end{document}